\documentstyle[11pt,hyfestconf,epsfig,graphics,twoside]{article}

\def \etal {{\em et al.}}
\def \eg {{\em e.g.}}

\def \cf {{\em c.f.}}

\begin{document}

\title{X-ray Observations of Groups and Clusters  -- A Relic of the
High Redshift Universe} 

\author{Richard Mushotzky}
\affil{NASA/Goddard Space Flight Center, Greenbelt, MD 20771\\
{\tt email: mushotzky@lheavx.gsfc.nasa.gov}}

\begin{abstract}
I review the evidence from clusters and groups of galaxies for `relic'
evidence of the high redshift universe. Contrary to the received wisdom,
clusters are old.  Their x-ray emitting intergalactic medium in massive
clusters is a reservoir of metal and energy injection from the main
epoch of star formation. There are strong indications that
non-gravitational energy was extremely important in the formation of
low-mass clusters and groups, implying that gravity did not dominate the
process on all scales below that of massive, rich clusters. The data
indicate that most of the metals in the universe are not in stars but in
the hot gas and thus the use of stars to trace metal formation and
evolution misses most of the processed material. The group data shows
that galaxies are very poor tracers of mass on the scale of
$10^{13}\,M_{\odot}$ and since groups are the `average' place in the
universe this probably holds for most of the visible universe.
\end{abstract}

\keywords{X-rays: galaxies --- galaxies: clusters --- cooling flows ---
galaxies: abundances}

\section{Introduction}
While classically clusters are presumed to be rather young (Gunn 1977),
and the present epoch is thought to be the era of cluster formation,
recent results strongly indicate that clusters are `old' and thus are
relics of the high-redshift universe. These indications are primarily
due to Rosat and ASCA results for $z<0.8$ clusters: the lack of
evolution in the x-ray luminosity function of clusters at $z<0.8$ (Jones
\etal\ 1998; Ebeling \etal\ 1999; Vikhlinin \etal\ 1998); the relatively
small change in the cluster temperature function out to $z\approx 0.8$
(Henry 1997; Donahue \& Voit 1999); the lack of evolution in the x-ray
luminosity temperature relation to $z\approx 0.5$ (Mushotzky \& Scharf
1997; Donahue \etal\ 1999); the lack of evolution in the metallicity in
the cluster gas to $z\approx 0.8$ (Fig.~1; Mushotzky \& Loewenstein
1997; Donahue \etal\ 1999); and the lack of evolution in the galaxy
velocity dispersion temperature relation (Tran \etal\ 1999). In addition,
the detection of massive clusters at $z\approx 0.8$ (Bahcall \& Fan
1998) and their possible existence at $z>1.2$ (Dickinson \etal\ 1999;
Stanford \etal\ 1997) are highly unlikely in a high-$\Omega$ universe
(however for a different opinion see Blanchard \etal\ 1999). Other,
independent, evidence is derived from HST and Keck observations that
indicate that the stars in cluster ellipticals are very old (Renzini
1999). However it is not known, at present, if the galaxies themselves
are old.  Thus for the high-redshift universe the question is: what can
we learn about the formation of large scale structure, galaxies and
clusters from studies of these relics?

\section{Cluster Results}
Clusters are, of course, the largest bound objects in the universe and
are thought to be fair samples of the content of the universe (baryons,
dark matter and metals).  As has been known for many years, x-ray
observations of groups and clusters indicate that most of the observable
baryons in these systems reside in the hot phase ($T>10^{7}$\,K,
Fig.~2). If clusters and groups are `representative' samples, the
stellar content of galaxies is, on average, $\approx 20$\% of the gas
mass (see Nevalainen 1999 for a recent detailed analysis). The mean
stellar `metallicity' (Fe) is $\approx 0.4$ `solar' (normalizing to the
solar abundances of Anders \& Grevesse 1989 and using the Jorgensen
\etal\ 1999 data, averaging out over the metallicity gradients and the
dependence of metallicity on luminosity). The metallicity of the gas is
$\approx 30$\% of solar (Fig.~3). Thus with 80\% of the mass in the gas
and a similar metallicity for the gas and the stars,
$\approx\threequarters$ of the metals are in the gas phase. Since there
are only subtle differences in the stellar spectra of `bulges' in groups
and clusters from the field, these results should apply to most galaxies
and therefore we expect that $\approx 20$\% of the observable baryons
and `most' of the total mass in stars is in bulges and that the rest of
the baryons are in the hot gas phase. It is amusing that this radical
result is actually found in semi-analytic models (Somerville \& Primack
1999).

What do these results imply for the formation and evolution of
galaxies?
\begin{itemize}
\item{Galaxies
are open systems -- much of their evolution involves outflow and inflow
of gas.} 
\item{The presently-observable stars do not contain most of the
metals.}
\item{One cannot study the history of metal creation by looking
only at the `relic' stars.}
\item{The relic gas contains most of the `stuff'
-- where is this gas when it is not in groups and clusters?}
\item{Estimates of the total metals created by counting only stars are
wrong --
\cf, the controversy (Madau this symposium) over the evolution of the
star formation rate in the universe.}
\item{There has been a lot more
metal creation/star formation in the universe than indicated by
normalizing to present day stellar data (Loewenstein \& Mushotzky
1996) because most of the metals are in the hot gas phase and not in
the stars. This is also essential for the semi-analytic models, which
require a large contribution from supernova energy to explain the
observed pattern of mass-to-light ratio and metallicity in
galaxies (see next section).}
\end{itemize}

\begin{figure}
\resizebox{0.9\textwidth}{!}{\rotatebox{270}{\includegraphics{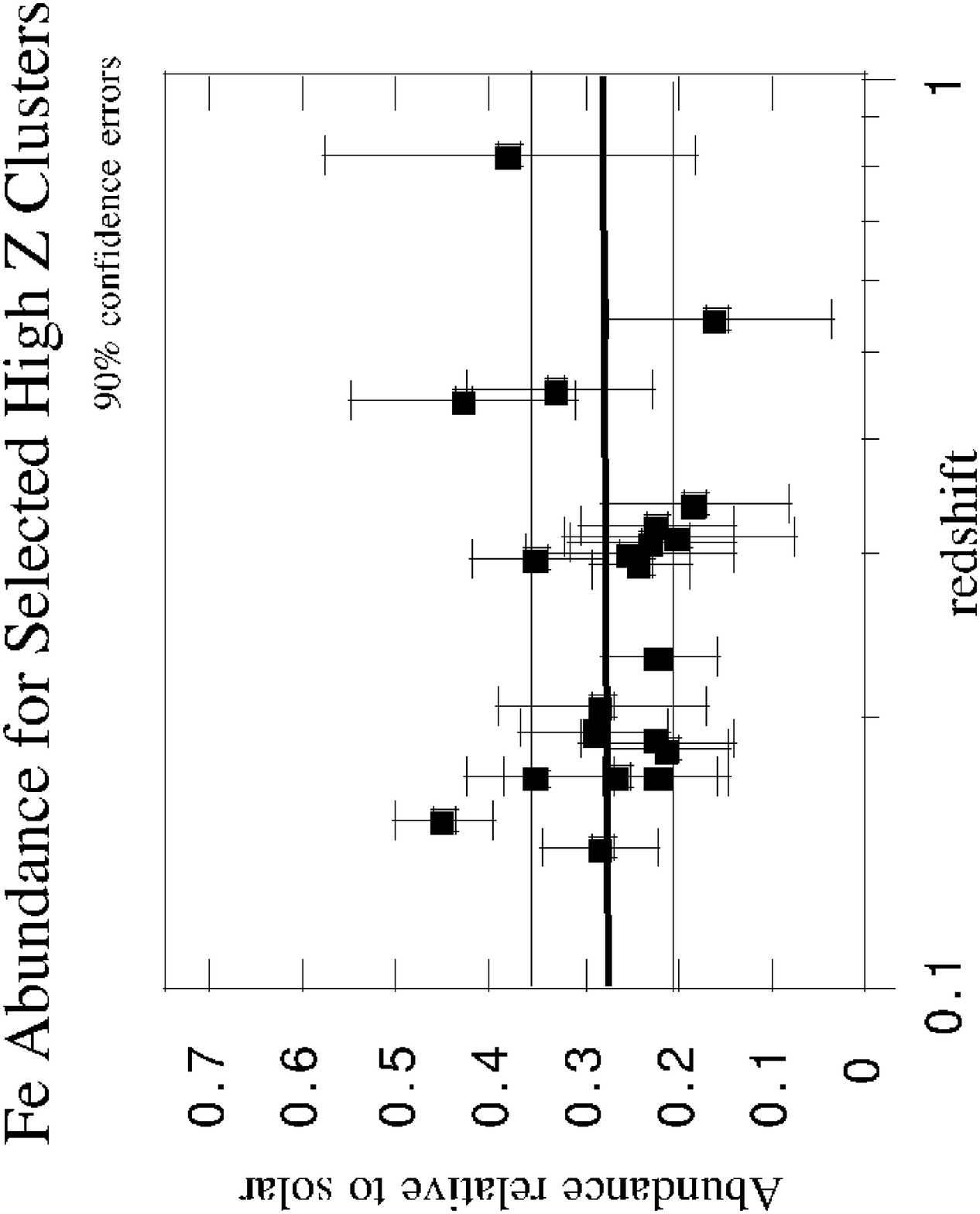}}}
\caption{Abundance {\em vs.\ }redshift of selected clusters of galaxies as
observed by ASCA. The
objects are selected to have the smallest uncertainties in abundance,
except for the point at
$z\approx 0.83$ which is the average of 2 clusters. The abundance scale is that of
Anders \& Grevesse
(1989). The horizontal lines are the $\pm 1\,\sigma$ range for the low-$z$ sample
shown in Fig.~3.}
\end{figure}

\section{The History of Star Formation has Very Strong Implications for the
Evolution of the Gas}

\subsection{Evidence for Extra Heat}
The gas in clusters and groups contains a history of the total energy
deposited in it. This energy originates from two general sources: (1)
gravity (infall shocks and mergers) and (2) star formation (supernova
and stellar winds).  Since massive star formation produces most of the
metals, we can use the metallicity as an estimate of the total heat
produced. For a fixed metallicity the fraction of the energy from star
formation increases as the mass of the system is reduced; in other
words the available energy per particle from infall is reduced as the
mass of the system is lowered while, for a fixed metallicity, the
energy produced by stellar processes remains constant. However it is
important to know when and where the energy is produced. It is
believed that for massive clusters most of the gravitational energy is
produced by the infall shock during ``cluster formation''. Theoretical
N-body hydrodynamical estimates of the creation of entropy (heat) via
shocks (Eke, Navarro \& Frenk 1998) are in good agreement with the
data for {\em massive} clusters.

However at low masses (Ponman, Cannon \& Navarro 1999) there is excess
entropy compared to these simulations. This effect has also been noted
by many in the context of the luminosity temperature relation and the
distribution of entropy in clusters (David \etal\ 1996). The
theoretical work predicts $L_{X}\propto T^{2}$ while the data show
that $L_{X}\propto T^{3}$ (Markevitchl 1989) and this can also be
explained by `extra' (\eg, non-gravitational) heat input (Kaiser
1991).

Detailed calculations (Loewenstein 1999; Tozzi \& Norman (TN) 1999;
Suginohara \& Ostriker 1998; Balogh \etal\ 1999) indicate that this
energy can be provided by star formation but requires a fairly high
efficiency of conversion of stellar energy into heat since the
$1-3$\,keV/particle that is required is about the maximum that can be
provided by the supenova that produce the metals. Since entropy
$\propto T/\rho^{\frac{2}{3}}$ (at the time of energy injection) for a
given heat input one gets more entropy if the density is low at time
of energy injection. These calculations strongly constrain the total
amount of energy and the epoch at which this energy is
injected. Following the scenarios of TN, where the entropy is
produced at a single epoch before the collapse of cluster, the mass
scale of the transition from a shock (gravity dominated) entropy to a
adiabatic model at a given mass scale is a strong function of the
total entropy and the redshift of energy injection. The appearance of
a floor in the entropy at a mass scale of $M\sim 10^{14}\,M_{\odot}$
($T\sim 1$\,keV) requires in their model a redshift of $z\sim 3-5$ for
the injection of the energy.  Suginohara
\& Ostriker (1998), who use a rather different set of constraints based
on the absence of thermal cooling runaway in clusters also require a
similar redshift for energy injection . While the situation is unclear
it seems as if certain structure formation scenarios (the
semi-analytic models presented by Valageas \& Silk 1999 and Wu, Fabian
\& Nulsen 1999) are not consistent with the required amount of extra
energy being provided solely by supernova.

\begin{figure}
\resizebox{0.9\textwidth}{!}{\includegraphics*[130,370][445,685]{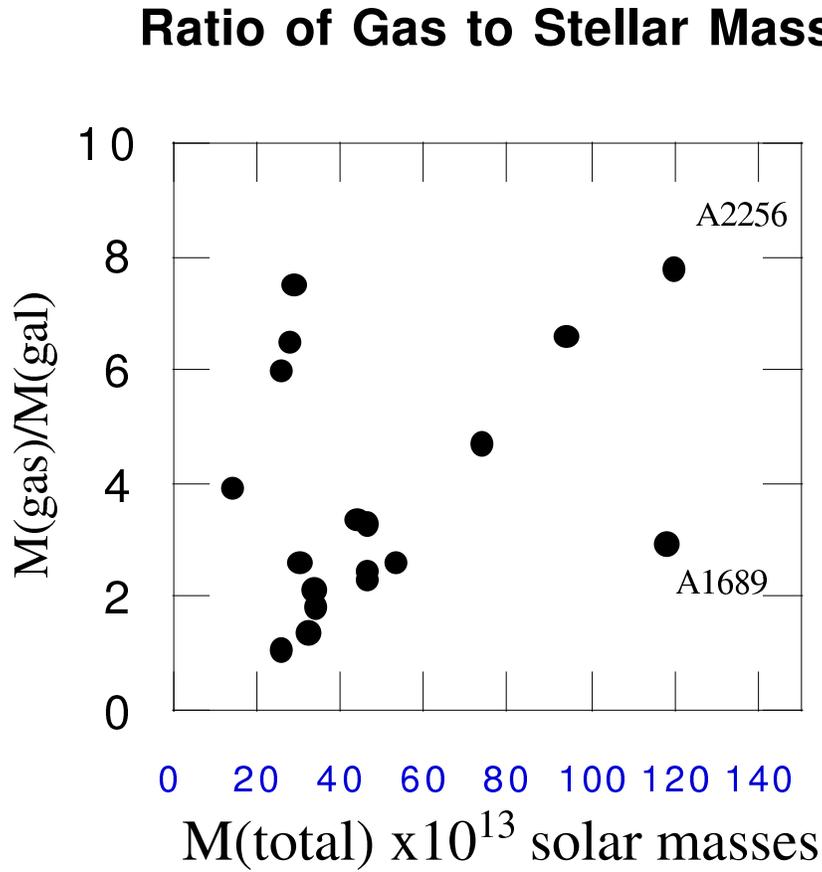}}
\caption{The ratio of gas mass to galaxy mass for a set of clusters
{\em vs.\ }total mass. The objects are selected from the literature
and much of the variance is due to different assumptions about the
conversion of light to mass, different radii analyzed for galaxy data
and x-ray images. All of these objects are at $z<0.1$. Recent work
(Schindler 1999) indicates that this does not vary much with
redshift.}
\end{figure}

\begin{figure}
\resizebox{0.9\textwidth}{!}{\includegraphics*[110,370][475,710]{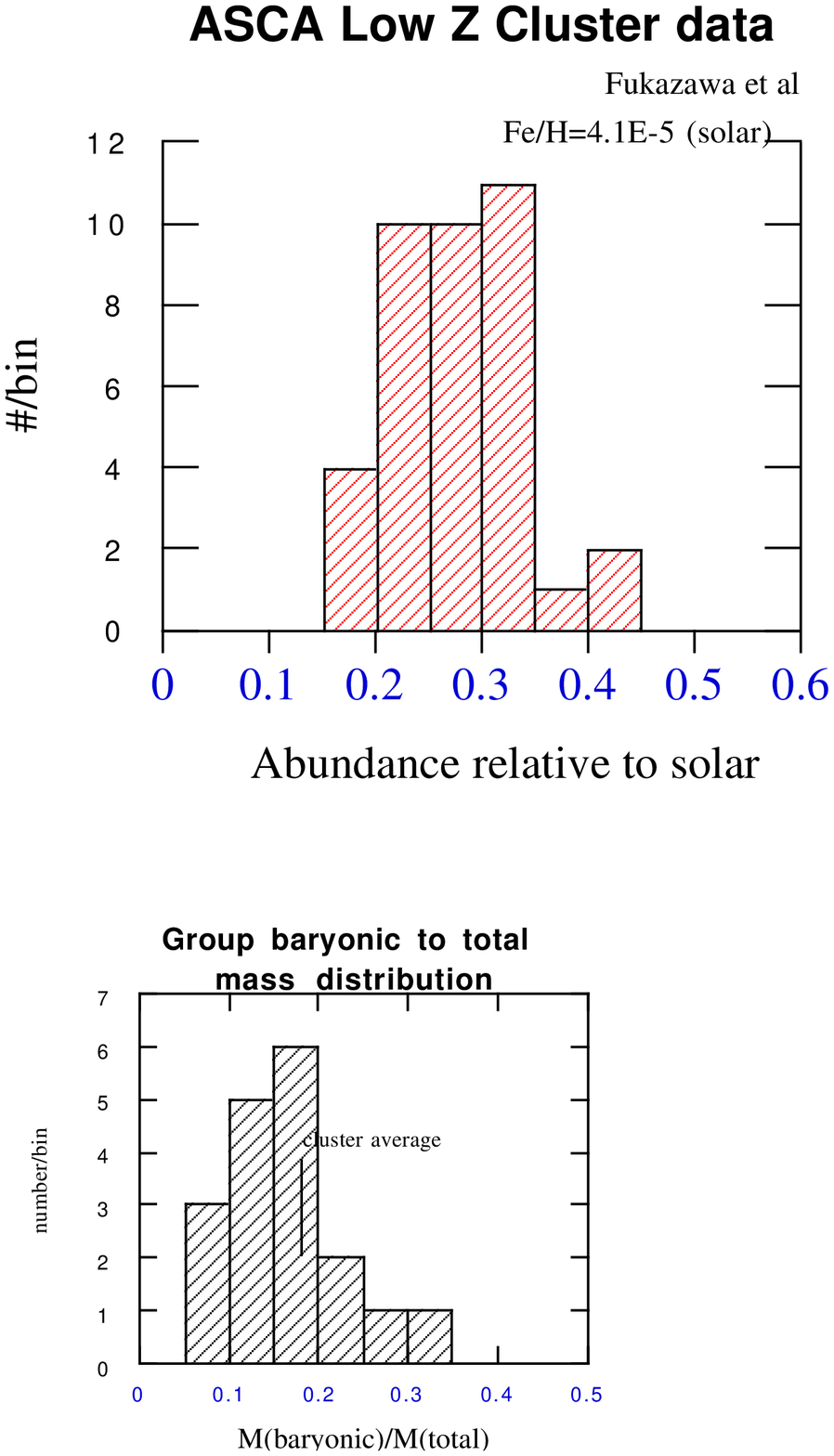}}
\caption{Metallicity of the gas for $z<0.1$ clusters. Most of the data are
drawn from Fukazawa (1996). The range in abundance is real and not an
aritfact  of large errors. A
typical error is $\pm 0.04$ in abundance. The vertical axis is the
number of clusters in each bin of abundance=0.05}
\end{figure}

This additional energy can also be checked (Loewenstein 1999; Davis
\etal\ 1999) by:
\begin{itemize}
\item{calculating the cluster mass {\em vs.\ }$T$ relation (Horner
\etal\ 1999) -- some of the `extra' energy goes into `temperature' and
at low masses the observed temperatures are ``too high'' for the the
N-body hydrodynamical model masses.  }
\item{measuring the surface
brightness distribution: the gas tends to get puffed-up and have a
flatter surface brightness distribution than predicted by the N-body
hydrodynamical models.}
\item{comparing the ratio of gas to total mass
{\em vs.\ }temperature. If there has been no energy input other than
gravity this should be roughly constant from object to object (White
\etal\ 1993).  However recent analysis (Mohr \etal\ 1999; Fujita \&
Takahara 1999) find that the ratio of gas mass to total mass rises as
$M_{\rm gas}/M_{\rm tot}\propto M_{\rm vir}^{0.4}$, that is, more
massive clusters have a larger fraction of their virial mass in gas
(as originally suggested by David \etal\ 1995 and as indicated in
Fig.~2).The obvious solution is that some of the gas gets expelled
in low mass systems.  This is further supported by the metallicity
found in groups and the wide range of other parameters found in groups
(see next section).}
\end{itemize}

\subsection{What can we say about when the heating occurred?}
The observations that low redshift elliptical galaxies are old and
have mostly stopped forming stars at $z<1$, the lack of evolution in
the cluster metallicity since $z\approx 0.8$ (Fig.~1) and the
predominance of ellipticals in rich clusters all indicate that most of
the star formation in clusters, and thus the source of the additional
entropy occurred at $z>1$. If the heating occurred only prior to
collapse (Ponman \etal\ 1999) one expects that low mass systems are
isentropic (Balogh \etal\ 1999) and thus have a steep temperature
gradient. {\em This has not been observed and the best data to date
indicate that groups are roughly isothermal.}

However, as noted above, the effects of heating are much more
efficient if they occur at early times, when the gas was of lower
density. It thus seems likely that the heat was created at moderate to
high redshift sometime near the collapse of the cluster.  Clearly,
detailed calculations are necessary to determine the critical
observations which can determine the amount of energy injected, the
epoch at which it is injected and the source of the energy.

While clusters may be `rare' high density perturbations, groups are
the average place in the universe (see next section). The effects of
heating are most pronounced in groups, as evidenced by their high
specific entropy, their relatively low gas fractions and flat surface
brightness profiles. Thus {\bf most} galaxies have been subjected to
this process. As indicated in the semi-analytic models heating has
profound implications on galaxy formation and the epoch of formation
-- the lesson that one learns is that ONE CANNOT UNDERSTAND GALAXY
FORMATION BY MODELING DARK MATTER HALOS ALONE . There are many other
implications of this; for example: when overdense regions ``get hot''
due to gravitational collapse (cluster and group formation), star
formation stops because the cooling times increase drastically (Cen \&
Ostriker 1999) -- this gives a time dependent bias such that overdense
regions which collapse early (rich clusters) cease star formation at
high redshifts and the regions in which star formation proceeds occur
in lower and lower overdensities at lower redshifts. The effects of
heating due to star formation has a strong feedback loop (Menci \&
Cavaliere1999) and strongly effects the timescale in semi-analytic
models.

If the true energy is $1-3$\,keV/particle, as indicated by the entropy
models (Loewenstein
\& Mushotzky; Loewenstein 1999) then the mean energy per particle of
most of the baryons exceeds the binding energy of even the most
massive galaxies and galactic winds were important for ALL galaxies
(not just dwarfs). {\em The x-ray data for groups and clusters strongly
argues that the galaxies are poor tracers of mass , baryons and metals
and use of them as tracers may give a very poor `picture' of the
universe.}

\section{Data on Cluster and Group Metallicity}
The Fe abundances are strongly peaked near $<Fe>\approx 0.3$ solar and
do not vary with redshift out to at least $z\approx 0.4$ and perhaps
$z\approx 0.8$ (Donahue \etal\ 1999). This lack of evolution of gas
metallicity combined with the absence of major star formation in the
same clusters strongly constrains the epoch of major metal
formation. The absence of large number of A stars means that the
period of massive star formation has ceased in these systems $\approx
2$\,Gyrs (the age of A stars) before the age of the objects at
$z=0.8$. For $\Omega = 0.2$, $\Lambda_{0}=0.8$, $H_{0}=65\,{\rm
km\,s^{-1}\,Mpc^{-1}}$ this implies that the epoch of metal formation
is these clusters is at $z>1.2$.  The highest-$z$ cluster with
measured Fe abundance is at $z\approx 1$ (Hattori \etal\ 1997) and if
we adopt this as the highest redshift at which cluster metallicity is
not evolving this would constrain the epoch of metal formation in
clusters to $z> 1.4$.

\subsection{Abundance Pattern:}
The data are fairly robust with respect to Fe and Si (Fukazawa \etal\
1998) moderate for S but poor for Mg, O, Ca, Ar, Ne.  Recent analysis
of ASCA, XTE and SAX data (Dupke \& White 1999) give a strong
indication that Ni maybe be very overabundant with respect to Fe.  One
needs to combine these patterns with supernova models to derive
constraints (Gibson, Loewenstein \& Mushotzky 1998) and the relative
numbers of type~I and type~II supernova. Unfortunately the
uncertainties in yields and SN models make strong statements very
uncertain.  The existence of abundance gradients in Si/Fe in about
$\onethird$ of all clusters and groups (Finoguenov \& Ponman 1999;
Ezawa \etal\ 1997) seem to show that the centers of clusters are
relatively enriched in type I products and the outer regions are
almost `pure' type~IIs. The bulk of all the heavy elements is in the
outer regions. As shown in Lowenstein \& Mushotzky (1996) the toal
mass of Si seen in clusters indicates that type~II's have dominated
the total metal production, consistent with the early origin of the
metals in bursts of star formation. However the situation is not yet
fully resolved (\cf, Arimtoto \etal\ 1997) and in order to determine
the true ratio of type~I to type~II will require better data for O, Ni
and Mg -- the elements that most sensitive to the relative number of
different types of supernova.

\section{Results For Groups}
Groups are the average place in the universe, over 70\% of all the
galaxies in the local volume of space occur in groupings and it is
very rare indeed to find truly isolated galaxies (Burstein 1999 pc. --
of the 2700 galaxies in the Tully Nearby Galaxies Atlas, only 35 are
not in a group, and none of those 35 are more than 0.75\,Mpc from
another giant galaxy). Thus the results for groups should apply to
{\em the universe as a whole if the x-ray emittings groups are
representative.}

Recent x-ray results show that large x-ray halos are a common property
of bulge dominated groups (Mulchaey \etal\ 1998) but are very rare in
spiral dominated systems.  The masses determined for the groups are
determined via the same techniques as for clusters (via use of
hydrostatic equilibrium and spatially resolved temperatures).

The derived masses are `typically' $\sim 3\times 10^{13}\,M_{\odot}$
at the viral radius (but only two groups have been measured to the
virial radius and if the systems are really dominated by
non-gravitational heat the extrapolation to the virial radius is
uncertain -- see Loewenstein 1999).

The space density of these systems is large and in rough agreement
with that predicted for virialized systems in LCDM models. Thus use of
an x-ray selected survey can derive the mass density of virialized
systems over a mass range $>100$ (from the lowest mass groups to the
richest clusters).

There is a very poor correlation between the optical richness of the
group and the x-ray temperature or luminosity (Mulchaey \etal\
1996). However there is an excellent agreement between the velocity
dispersion as derived from observations of the numerous small galaxies
in x-ray luminous groups (Mulchaey \& Zabludoff 1998) and the x-ray
temperature. This essentially implies that in groups the optical light
does not trace mass at all! The x-ray emission is a much more direct
tracer of mass than optical light and determination of the group mass
and mass distribution will be a strong test of all structure formation
models. This is a serious warning about the use of optical light to
trace mass in the universe since most galaxies are in poor groups.

The baryonic fraction, the gas to stellar mass and the $M/L$ ratio all
show wide variations in groups, but the mean values are distributed
around the cluster average (Fig.~4)

\begin{figure}
\resizebox{0.9\textwidth}{!}{\includegraphics*[120,75][380,310]{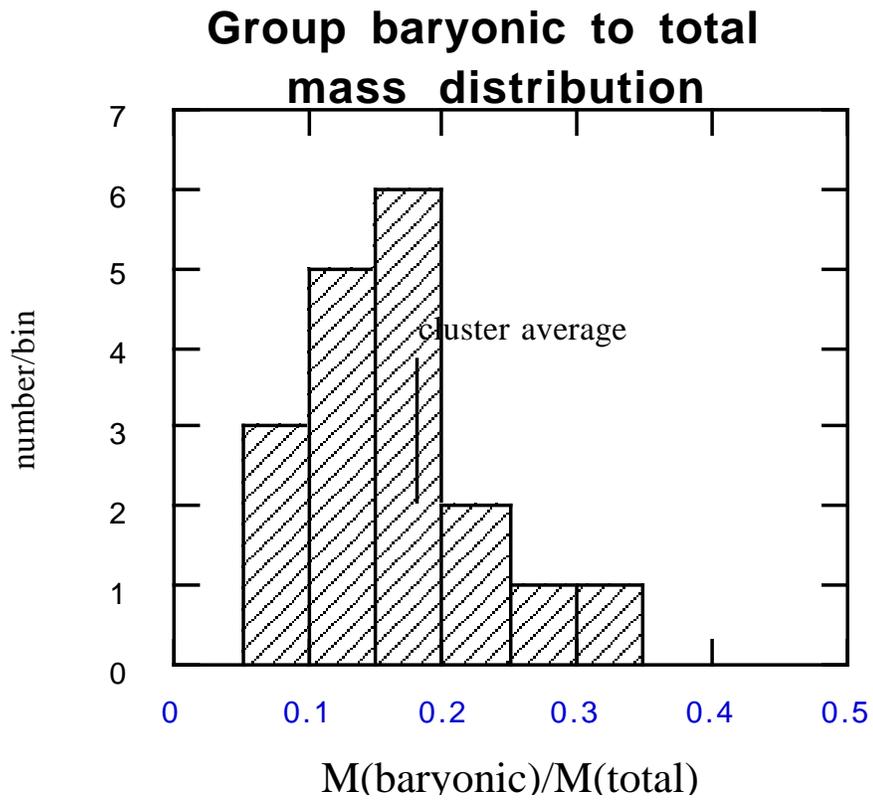}}
\caption{Ratio of baryonic (gas + stars) to total mass for groups. The
cluster average is
indicated with a vertical bar. The range for clusters is considerably
smaller than in groups.}
\end{figure}

\subsection{Abundance Data}
The metallicity in groups has a much larger range than that found in
clusters and is often less
(Fig.~5).

\begin{figure}
\resizebox{0.9\textwidth}{!}{\includegraphics*[100,440][440,690]{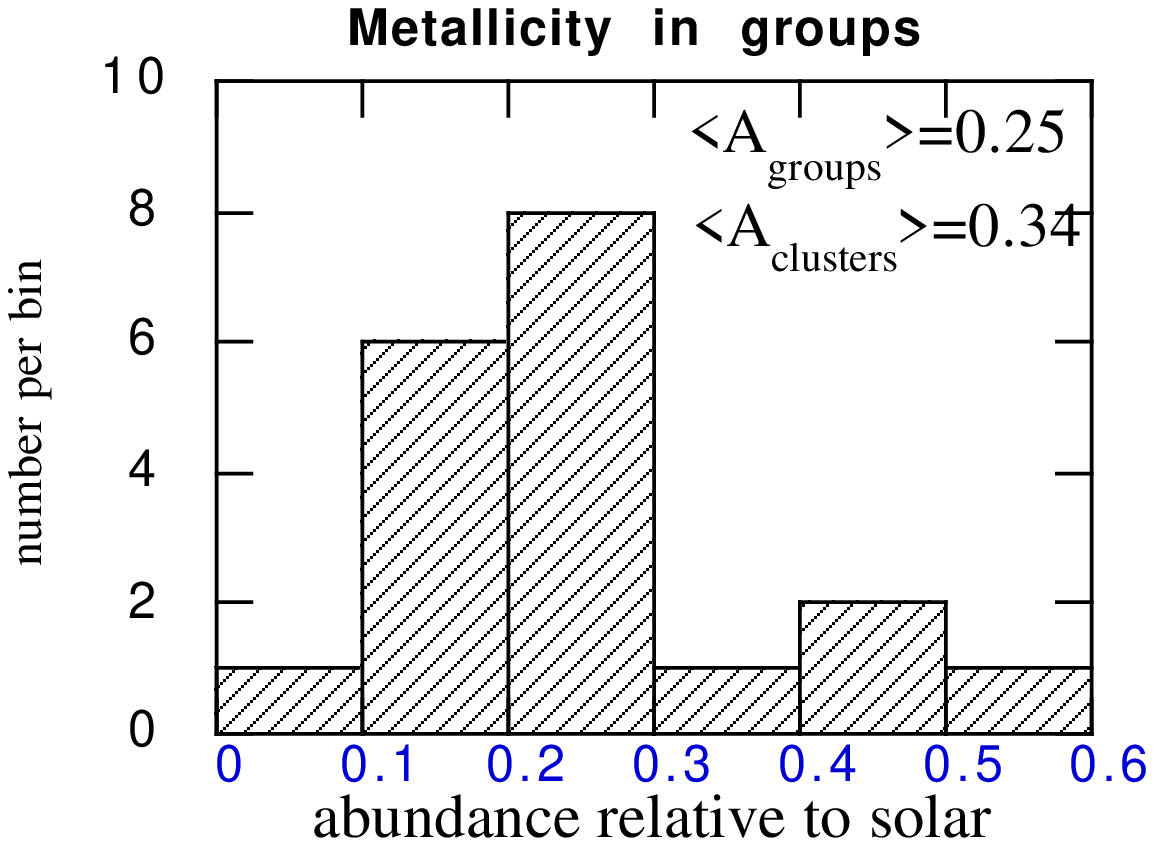}}
\caption{The abundance in groups. Compare to Fig.~3 for rich clusters
and note the much larger spread.}
\end{figure}

The ``iron mass to light ratio'' (ratio of total mass in Fe
(stars+gas) to the total light -- Renzini \etal\ 1997) tends to be
much lower than in clusters. If we believe that the stars visible
today in the groups are the tracers of the population that produced
the metals then this indicates that the groups have lost a substantial
fraction of their metals.

The abundance pattern is not well determined (Davis \etal\ 1999) but
the Si/Fe ratio is smaller than in clusters (Fukazawa \etal\ 1998)
indicative of either a larger ratio of type~I to type~II supernova
than in clusters, which is hard to understand, or a loss of type~II
material (Davis \etal\ 1999). Since the groups tend to have a higher
specific entropy than clusters (normalized to the expected shock
value) and have a wider range in baryon fraction, all these signatures
are consistent with mass ejection via the energy input from stellar
sources associated with massive star formation (SN+winds) on the group
mass scale.

\subsection{What does this mean for the high redshift Universe?}
The low iron mass to light ratios strongly constrain evolution
scenarios and indicate (1) either these systems have lost much of
their baryons and metals (winds from galaxies and groups) and/or (2)
they have been much more strongly effected by infall of primordial
material than clusters. {\em And thus even the most massive ``normal
systems'' have been affected by non-gravitational processes.}

\section{Conclusions:}
The picture of the high redshift universe that one obtains from x-ray
observations of groups and clusters in the $z<0.8$ universe is
radically different from what was the standard picture five years
ago. Clusters are old, and because they can hold onto all their
material (and even accrete from the IGM), they are the only systems
that are representative of the universe as a whole.  There are strong
indications from observations of groups and clusters that
non-gravitational effects are extremely important in the formation of
galaxies and groups and that gravity is not dominant on all
scales. One finds that at the group scale that galaxies are very poor
tracers of mass. The combination of all these effects may help to
explain the difficulty of comparing numerical N-body simulations of
structure formation against the available galaxy data. Finally the
importance of feed-back from non-gravitational energy sources, as
indicated in the semi-analytic models, cannot be underestimated. We
anticipate that the study of large scale structure with x-ray samples
will radically change our understanding of the universe.

In the next year, with the new data from Chandra (now operating), XMM
and Astro-E there will be a qualitative change in the signal to noise,
angular resolution and spectral resolution of the data and many of the
conclusions reached in this review will be strongly tested.

\acknowledgements
I would like to thank the organizers for the chance to present the
``Hy-Energy'' view of the
universe and for an exciting and well organized meeting. I would like to
thank my
collaborators: M Loewenstein, U. Hwang, J. Mulchaey, D. Davis, C. Scharf
and D. Horner
who have done so much of this work. Andrew Bunker is thanked for his
assistance in assembling this manuscript.


\end{document}